\documentclass[twoside]{article}

\usepackage{titling} 
\usepackage{hyperref} 
\usepackage{authblk}
\usepackage[T1]{fontenc}
\usepackage[utf8]{inputenc}
\usepackage{slashed}
\usepackage{dsfont}
\usepackage{amsfonts}
\usepackage{amsmath}
\usepackage{cleveref}

\usepackage{blindtext} 
\usepackage[dvipsnames]{xcolor}
\usepackage[sc]{mathpazo} 
\usepackage[T1]{fontenc} 
\usepackage{microtype} 

\usepackage[english]{babel} 

\usepackage[hmarginratio=1:1,top=32mm,columnsep=20pt]{geometry} 
\usepackage[hang, small,labelfont=bf,up,textfont=it,up]{caption} 
\usepackage{booktabs} 

\usepackage{lettrine} 

\usepackage{enumitem} 
\setlist[itemize]{noitemsep} 

\usepackage{abstract} 

\usepackage{titlesec} 
\renewcommand\thesection{\Roman{section}} 
\renewcommand\thesubsection{\roman{subsection}} 
\titleformat{\section}[block]{\large\scshape\centering}{\thesection.}{1em}{} 
\titleformat{\subsection}[block]{\large}{\thesubsection.}{1em}{} 

\usepackage{fancyhdr} 
\pretitle{\begin{center}\Huge\bfseries} 
\posttitle{\end{center}} 
\title{Vacuum Polarization of Generalized Quantum Electrodynamics in the Heisenberg Picture} 
\author[1]{David Montenegro\thanks{dmonte@ift.unesp.br}}
\affil[1]{S\~{a}o Paulo State University (UNESP), Institute for Theoretical Physics (IFT), R. Dr. Bento Teobaldo Ferraz 271 CEP 01140-070, S\~{a}o Paulo, SP}
\date{} 
\renewcommand{\maketitlehookd}{%
\begin{abstract}
\noindent In this work, we consider the Generalized Quantum Electrodynamics proposed by Podolsky in Heisenberg picture via Källén methodology. We investigate the effects of higher order derivatives to understand the qualitative and quantitative aspects of vacuum polarization. In addition, the most general structure of induced current and polarization tensor that emerge naturally by a perturbative scheme "à la" Källén are also obtained. Afterward, we discuss the physical implication of charge renormalization in the perspective of unitary and stable Podolsky theory and verify the influences of Podolsky's length on vacuum polarization.  
\end{abstract}}

\begin{document}

\maketitle


\section{Introduction}
\label{intro}

 
First-order Lagrangians involve important properties and behavior of matter field. In many cases, this setting is suitable for an appropriate physical description in different areas of physics. It would be interesting to generalize for other contexts in which the appearance of higher derivative Lagrangians ( terms with derivatives of second-order or higher in the Lagrangian) play a fundamental rule and give rise to new phenomena where low-order Lagrangians is dismissed. A long-time at 1850, Ostrogradski developed a procedure to deal with non-degenerate higher derivative systems \cite{Ostrogradski}, despite being mathematically self-consistent, it exhibited a set of unwelcome features as instability, negative energy modes, "ghost-particle" and loss of unitarity \cite{pais}. Even though these aspects still impose a systematic deficiency when higher derivative theories interact with a stable Hamiltonian system, higher derivative Lagrangians continue motivating further investigation due to useful insight into underlying aspects of the theory.

In a more general context, higher derivatives are very informative and lead us towards a better understanding of theory. For example, in the quantum regime, it may improve the convergence of propagator when high frequency controls loops diagrams behavior and gets rid of all ultraviolet (UV) divergence. Following this line of thought, extended gravity models have gained theoretical interest when a simple extension of Einstein-Hilbert action lays out perturbatively a renormalizable and asymptotically free theory \cite{Stelle}. Additionally, in three-dimensional gravity models, also, show higher derivatives suppress UV divergence integral as a natural regulator \cite{Deser}. Another important application is to further explore string theory where such procedure is used for evaluating the world-sheet by a toy model that gives rise to different inflationary systems at Plank scale \cite{String} and, in supersymmetry theory, the scalar supermultiplet fields govern the construction of supergravity models, for more details see \cite{supersymmetric}. Although Ostrogradski's instability offers a typical challenge to lead with higher derivatives \cite{Lu}, It is worth pointing out the application of such Lagrangians has outstanding properties by the theoretical point of view and generalized applicability.

As the purpose of this work, we study the formulation of higher derivatives on ordinary quantum electrodynamics ($QED_4$). This extension, by adding a second-order derivative gauge term, is known as Generalized Quantum Electrodynamics ($GQED_4$) and provides a richer theoretical scenario with a new number of degrees of freedom and families of solution. Initially, It was suggested by Bopp \cite{Bopp} and Podolsky \cite{Podolskyorigin} in 1942 to solve pathologies from $QED_4$, for example, the static self-energy of point charge \cite{podsel}, the old classical "$4/3$ problem" solved by Frenkel \cite{43problem}, and infinity at 1-loop correction from meson field theory removed by Green \cite{Greenpod}. Furthermore, the discrepancy found in $QED_4$ experiments such as the magnetic moment of muon \cite{hydro} and $1S$ ground-state Lamb shift in atomic hydrogen \cite{bom,muon} open a new theoretical insight over modification in standard electrodynamics. Therefore, one can also mention from a higher derivative theories perspective that ordinary electromagnetism was incomplete and not well accurate.

Besides that, a very useful feature of $GQED_4$ is the equivalence, apart from a surface term, to other different versions of second-order gauge theories that preserve gauge invariance and linearity \cite{Cuzinattogauge}. Moreover, a basic and technical question remains concerning gauge invariance of higher derivative theories since an appropriate gauge condition will reflect on the internal structure of quantum fluctuation. In other words, we must settle out the gauge fixing problem to get rid of undesirable non-physical degrees of freedom. For answering this question, it was proved by \cite{nomixing1}, the no-mixing gauge drawn the proper solution for the correct degree of freedom.

As it is well known, reasonable physical theory demands the absence of Ostrogradski problems. The standard Noether theorem provides a sufficient but not necessary condition for detecting such instability and a new technique based on the Lagrangian anchor showed stable dynamics due to another bounded conserved quantity \cite{russo}. According to this, It has been proved, in the previous reference, that Podolsky is dynamically stable for both the classical and quantum levels, and unitarity endowed from BRST \cite{Nogueira}. Hence, GQED not matching the class of effective field theory.

It is important to observe that we can accomplish a unitary equivalence representation between Interaction Picture (IP) and Heisenberg Picture (HP) in non-relativistic quantum mechanics, where there is a finite degrees of freedom, in other words, the expectation value for correspondent observables presents the same result in both pictures. However, this idea is misleading in quantum field theory, where infinite degrees of freedom are taken into account and Haag's theorem showed a mathematical inconsistency for IP \cite{Haag}. Also, this question is more serious when divergent quantities from Feynman diagrams are involved and the physical equivalence between HP and IP remains undermined. 


Källén developed a quantization program for the formulation of $QED_4$ in the HP \cite{Kallen1,Kallen2}. To tackle renormalization and regularization problems, He assumes a perturbative recourse in which the field operators in HP are given as an expansion into series in power of small coupling constant, and thus directly introducing into differential equations of motion. Other quantum systems where HP has been applied to the Thirring model \cite{Lunardi}, $QED$ in three dimensions \cite{Lunardi2}, and Scalar Quantum Electrodynamics (SQED) \cite{Beltran}. In the present work, we expect to obtain a radiative correction of self-energy photon up to the first-order loop by Källén's formalism on $GQED_4$.

This work is organized as follows. In section \ref{GQED}, we briefly discuss the generalized quantum electrodynamics of Podolsky. It is worthwhile an introduction of Dirac quantization in section \ref{DiracField}. The previous sections give support to implement the perturbative method of Källén in section \ref{Perturbation}. In section \ref{VAC}, we treat new characteristics of vacuum polarization which emerge from the Podolsky term. In our final remark in the last section \ref{conclu}, we shall establish the results and conclusion. Adopting the Einstein summation convention for indices and in all formulas we use the natural units, $\hbar = c = 1$ for the article.

\section{Generalized Quantum Electrodynamics schedule}\label{GQED}

Here, we shall give a brief review of Generalized Quantum Electrodynamics and introduce the essential aspects that will be explored in subsequent sections. Thus, starting with Podolsky Lagrangian density  

\begin{equation}\label{lpod}
 \mathcal{L} = -\frac{1}{4}F^{\mu\nu}F_{\mu\nu} + \frac{a^2}{2} \partial^\mu F_{\mu\beta}\partial^\alpha F_{\alpha\beta}
\end{equation}
 
where field strength tensor $ F^{\mu\nu}=\partial^{\mu}A^{\nu} - \partial^{\nu}A^{\mu}$. The new parameter $a$ has dimension of length and may define the Podolsky cutoff. The metric signature $x_4 = it$ and $ds^2 = - dx_\alpha dx^\alpha $. This Lagrangian is the only linear extension of quantum electrodynamics up to second-order derivative invariant under gauge $U(1)$ and lorentz group \cite{Cuzinattogauge}. The Bianchi identity reads
 
\begin{equation}
\partial_\alpha F^{\beta \gamma} + \partial_\gamma F^{\alpha \beta} + \partial_\beta F^{\alpha \gamma} = 0
\end{equation} 

and continues unchanged after the generalization of Podolsky. The Euler-Lagrange equations of motion for \eqref{lpod} gives

\begin{equation}\label{elpod}
(1 - a^2 \Box ) \partial_\mu F^{\mu \nu} = 0;
\end{equation}


where $\Box = \partial^\alpha \partial_\alpha$. Before discussing the details of equations of motion. We shall pay attention that all gauge theories must have a constraint to avoid excitation in Hilbert space of unphysical modes so that an appropriate gauge choice is necessary. The standard procedure in $QED_4$, which is widely explored, corresponds to the framework of Lorenz gauge $ \Omega_L [A] = \partial_\mu A^\mu$ \cite{Podolskyorigin}. Although this Lorentz invariant condition linear in $A^\mu$ appears a more reliable option, all redundant variables cannot be eliminated from \eqref{elpod}. In other words, longitudinal components remain in the dynamic due to the new degree of freedom in Lagrangian \eqref{lpod}. According to which, a new gauge transformation called generalized Lorentz condition $\Omega_G [A]=(1 - a^2\Box)\partial_\mu A^\mu $, as we guess by looking at \eqref{elpod}, was proposed in \cite{galvao}, one application see for instance \cite{bufalogauge}. However, a simple argument based on the fact that the insertion of $\Omega_G [A]$ must respect the same order derivative of the Lagrangian level. Then our guess fails to give the proper correction. To circumvent this problem, we must deform the generalized gauge conditions to $\Omega_P [A]=(\sqrt{ 1 - a^2 \Box }) \partial_\mu A^\mu $ \cite{nomixing1}, which is known as no-mixing gauge and suits the right order derivative of field equations.

Now, we are familiar with gauge symmetry, we would like to take a step towards a better understanding of Podolsky equations in the free case. The usual treatment enforces joining gauge constraint as a Lagrange multiplier $\xi$ into the Lagrangian density \eqref{lpod}. Thus  

\begin{equation}\label{lpod1}
\mathcal{L} = -\frac{1}{4}F^{\mu\nu}F_{\mu\nu} + \frac{a^2}{2} \partial^\mu F_{\mu\beta}\partial_\alpha F^{\alpha\beta} - \frac{1}{2 \xi}(1 - a^2 \Box) (\partial_\mu A^\mu)^2 
\end{equation}

with a suitable chosen of gauge $\xi=1$, the new equations of motion are given by   

\begin{equation}\label{1532}
    (1 - a^2 \Box ) \Box A^\mu (x) = 0;
\end{equation}

we may split up the potential vector over two physical different contributions.

\begin{equation}\label{super}
A^\mu (x) = A_{Max}^\mu(x) + A_{Pod}^\mu (x)  
\end{equation}

where Maxwell-Lorentz $A_{Max}^\mu(x)$ and Podolsky $A_{Pod}^\mu (x)$ fields coordinate satisfy the following equations

\begin{equation}\begin{aligned}\label{era}
(1 - a^2 \Box ) A^\mu (x) &= A_{Max}^\mu (x), \quad a^2 \Box A^\mu (x) =  A_{Pod}^\mu (x) \\
(1 - a^2 \Box ) A_{Pod}^\mu (x) &= 0, \quad \quad \quad   \quad
\Box A_{Max}^\mu (x) = 0 
\end{aligned}\end{equation}

We shall interpret $A^\mu (x)$ as sum of two photons: $A_{Max}^\mu (x)$  massless and transverse and $A_{Pod}^\mu (x)$ massive and longitudinal \cite{Podolskyorigin}. From adding a coupling between current $J^\mu (x)$ and $A_\mu (x) $ in the function \eqref{lpod}, one can derive the green function through these new equations of motion in Fourier space

\begin{eqnarray}\label{431}
\bigg[ (1 + a^2 p^2)  (-p^2 g^{\alpha\beta} + p^\alpha p^\beta ) - \frac{1}{\xi} (1 + a^2 p^2) (p^\alpha p^\beta ) \bigg] A_\alpha &=& J^\beta \\
\end{eqnarray}

The inverse of the operator in brackets reads

\begin{equation}
D(p^2)^{\mu \alpha} = \frac{g^{\mu \alpha}   - \frac{p^\mu p^\nu}{p^2}}{p^2(1 + a^2 p^2)} + \frac{\xi }{p^2 (1 + a^2 p^2)}\frac{p^\mu p^\nu}{p^2}
\end{equation}

and reflects the nature of \eqref{super}. If we take again $\xi = 1$, the green function of equation \eqref{1532} becomes

\begin{equation}\label{gfpod9001}
D_P (p^2) = \frac{g^{\mu \alpha}}{(1 + a^2p^2)p^2} 
\end{equation}

For later use, the retarded $D_P^R (p^2) = -\Theta(x_o) D_P(p^2)$  and advanced  $D_P^A (p^2) = \Theta(- x_o) D_P (p^2)$ green function are



\begin{subequations}\begin{align}
\label{retb}
D_R^{P} (x) &= \frac{1}{(2 \pi)^4} \int d^4 p e^{ipx} \bigg( \mathcal{P} \frac{1}{p^2} - \mathcal{P} \frac{1}{p^2 + a^{-2}} + i \pi (\delta(p^2) - \delta(p^2 + a^{-2})) \epsilon(p) \bigg), \\
 \label{avcb}
D_A^{P} (x) &= \frac{1}{(2 \pi)^4} \int d^4 p e^{ipx} \bigg( \mathcal{P} \frac{1}{p^2} - \mathcal{P} \frac{1}{p^2 + a^{-2}} - i \pi (\delta(p^2) - \delta(p^2 + a^{-2})) \epsilon(p) \bigg)
\end{align}\end{subequations}

respectively. Where principal value $\mathcal{P}$ and we define sgn function $\epsilon(p)$ and heaviside function $\Theta(x_o)$

\begin{equation}
    \epsilon (p) \equiv \frac{p_o}{|p_0|}, \quad \Theta (p) \equiv \frac{ 1 + \epsilon (p)}{2},
\end{equation}

respectively. We can define Podolsky mass as $ m_p \equiv \frac{1}{a}$. The propagator above relates the superposition principle made in \eqref{super}. One may write the general situation where the differential equation depends on the inhomogeneous part

\begin{equation}\label{wtf1}
    (1 - a^2 \Box )\Box A^\mu (x) = g(x)
\end{equation}

the resulting equations of motion reads 

\begin{equation}
    A_\mu (x) = A^{(0)}_\mu (x) + \int d^4 x' D^P_R ( x - x') A_\mu (x')
\end{equation}

where $A^{(0)}_\mu (x)$ is the homogeneous solution of differential equation \eqref{1532}. From the representation \eqref{retb} and \eqref{avcb}, the Green functions satisfy the inhomogeneous equations

\begin{equation}
    (1 - a^2 \Box )\Box D^P_R (x-x')  = (1 - a^2 \Box )\Box D^P_A (x-x') = - \delta^4 (x-x')
\end{equation}

where $\delta^4 (x)$ is the Dirac delta distribution.

\section{The Dirac Field}\label{DiracField}

Now, we overview the free Dirac formalism and seek to write green functions based on the Dirac operator field. Let us first analyze, the standard Dirac Lagrangian density

\begin{equation}\label{Diracfree}
\mathcal{L}_{\psi} =
\Bar{\psi}_a (x) ( (\gamma^{ab})_\mu \partial^\mu + \delta^{ab} m) \psi_b (x)     
\end{equation}

Where antifermion and fermion functions are $\Bar{\psi}_a (x) $ and  $\psi_a (x)$, respectively. The small Latin indices summed over $\in \{ 0,3 \}$ represent spinorial indices and $m$ the electron mass. The matrices $ (\gamma^{ab})_\mu $ obey Clifford algebra $\{ \gamma^\mu,\gamma^\nu \}_{ab} = 2 g^{\mu\nu} \delta_{ab}$. The equations of motion for a free electron can be obtained from Hamilton's principle of \eqref{Diracfree} and reads as

\begin{subequations}\begin{align}
\label{fdsw}
( (\gamma_\mu )^{ab}  \partial^\mu + \delta^{ab}  m ) \psi_b (x)=0, \\
\label{fdsw2}
\bar{\psi}_a (x)  ( ( \gamma_\mu )^{ab}  \overleftarrow{\partial}^\mu + \delta^{ab}  m ) =0
\end{align}\end{subequations}

The properties of the Dirac equation are extensively discussed in Ref. \cite{Bogoliubov}. Expanding the classical solutions of the free equations in plane waves 

\begin{subequations}\begin{align}
\label{DKP   solution 3}
\psi_a (x) &=\int\frac{d^3p}{(2\pi)^\frac{3}{2}}\sum^2_{r=1} a^r(\mathbf{p}) u_a^r(\mathbf{p})e^{-ipx} + \int\frac{d^3p}{(2\pi)^\frac{3}{2}}\sum^2_{r=1} b^{\dag r}(\mathbf{p})  v_a^r(\mathbf{p})e^{ipx}, \\
\overline{\psi}_a (x) &=\int\frac{d^3p}{(2\pi)^\frac{3}{2}} \sum^2_{r=1} a^{\dag r}(\mathbf{p})\bar{u}_a^{r}(\mathbf{p})e^{ipx}+\int\frac{d^3p}{(2\pi)^\frac{3}{2}}\sum^2_{r=1} b(\mathbf{p})\bar{v}_a^{r}(\mathbf{p})e^{-ipx}
\end{align}\end{subequations}

We shall list $\bar{\psi}(x) = \psi^*(x)\gamma^4$, and $u_a^{*r} u_a^s = \delta^{rs}$. Following Fermi-Dirac commutation relation of creation and annihilation operator at equal times  

\begin{equation}\label{comopcreani}\begin{cases}
[\hat{a}(\mathbf{p}),\hat{a}^\dag(\mathbf{p}')]=\delta(\mathbf{p}-\mathbf{p}' ),\\
[\hat{b}(\mathbf{p}),\hat{b}^\dag(\mathbf{p}')]=\delta(\mathbf{p}-\mathbf{p}' ).\\
\end{cases}\end{equation}

The next task is now to derive the fundamental commutation rules for arbitrary times through the ones at equal times. In advance, we shall consider  $\psi_a (x)$ as a Grassmannian operator rather than a Grassmannian function. The anticommutation of Dirac operator from \eqref{comopcreani} becomes 

\begin{equation}
\left\{ \Bar{\psi}_a(x), \psi_b (x') \right\} = -i S_{b a } (x'- x)  
\end{equation}

where 

\begin{equation}\label{spinin}
S_{a b }(x-x')  = \frac{-i}{(2\pi)^3 } \int d^4p e^{ip(x-x')}  (i \gamma p - m)_{b a}   \delta(p^2 + m^2) \epsilon(p)
\end{equation}

is the Jordan-Pauli distribution. The vacuum expectation value of the commutation of operator $\psi(x)$  

\begin{equation}\label{s11}
\langle 0 [\Bar{\psi}_a (x) , \psi_b (x')] 0 \rangle  = S^{(1)}_{b a} (x'- x) 
\end{equation}

where

\begin{equation}\begin{aligned} \label{s1}
S^{(1)}_{b a}(x - x')  &=  \frac{1}{(2\pi)^3} \int d^4p e^{ip(x-x')} (i \gamma p - m)_{b a}  \delta(p^2 + m^2)
\end{aligned}\end{equation}

The distribution function \eqref{spinin} and \eqref{s1} will be taken into account in perturbation calculation of field operator $\psi_a (x)$ in the next section. The retarded $S_{R} (x) = -\Theta(x_o) S(x)$  and advanced $S_A (x) = \Theta(-x_o)S(x)$ functions are

\begin{subequations}\begin{align}
\label{retf}
S_R(x) &= \frac{1}{(2\pi)^4}
\int d^4p e^{ixp} (ip\gamma - m)[P \frac{1}{p^{2}+m^2} + i \pi \epsilon(p) \delta(p^{2} + m^2) ]  \\
\label{avcf}
S_A(x) &= \frac{1}{(2\pi)^4}
\int d^4p e^{ixp}(i p \gamma - m)[P \frac{1}{p^{2}+m^2} - i \pi \epsilon(p) \delta(p^{2} + m^2) ] 
\end{align}\end{subequations}

Now we shall explore a straightforward generalization of homogeneous equation \eqref{fdsw} that leads us to the interacting system where dynamic can be represented by the inhomogeneous equation.

\begin{equation}\label{ihf}
(\gamma_\mu  \partial^\mu + m)\psi(x)= g(x)
\end{equation}

where $g(x)$ is the "source". Thus, we can address a large set of problems since non-interacting theories represent a slight set of them. By the formulation of Green's theorem with specified boundary conditions, we can find out the complete solution for the above equation

\begin{equation}
    \psi (x) = \psi^{(0)}(x) - \int d^4x' S_R (x-x') g(x') 
\end{equation}

Where $\psi^{(0)}(x)$ is the solution of homogeneous equation. For a more detailed explanation, see \cite{green}. In a similar way, the inhomogeneous conjugate of \eqref{ihf} reads

\begin{equation}\label{wtf2}
    \bar{\psi} (x)  ( ( \gamma_\mu )  \overleftarrow{\partial}^\mu + \delta  m ) = g(x)
\end{equation}

and the solution is

\begin{equation}
   \bar{\psi} (x) = \bar{\psi}^{(0)}(x) - \int  g(x') S_A (x'-x) dx'
\end{equation}

where $\bar{\psi}_a (x)$ satisfies \eqref{fdsw2}. Thus, 

\begin{equation}
( \gamma_\mu \partial^\mu + m ) S_R(x) = ( \gamma_\mu \partial^\mu + m ) S_A(x) = - \delta^4 (x)
\end{equation} 

By virtue of Noether's theorem,  charge conservation is associated with a global continuous symmetry U(1) through transformation law $ \psi'(x) = e^{i e \alpha} \psi(x) $, $\alpha$ being the gauge coupling. As a result, it gives origin to classical current density $j^\mu (x) = e \bar{\psi}(x) \gamma^\mu \psi(x)$, and its connection to operator level is constructed by

\begin{equation}\label{CORRP}
j^{\mu} (x) \equiv \frac{ie}{2}[\Bar{\psi}(x), \gamma_\mu \psi(x)].
\end{equation}

It is possible to check this current operator matches the normal order $ j^{\mu} (x) = i e :\Bar{\psi}(x) \gamma_\mu \psi(x): $ \cite{Kallen1}. Then, the main consequence of this prescription reads     
 
\begin{equation}\label{4corrienteq2} 
\langle0|j^\mu(x)|0\rangle = 0,
\end{equation}

Thereby, the symmetrization in \eqref{CORRP} yields a redefinition of the current operator in terms of normal form which guarantees the vacuum expectation value vanishes. So far, we review the basic stuff of free theory from fermion and boson. In the next sections, we will investigate a perturbation scheme to analyze the quantum correction of an interacting system in HP.

\section{Perturbation $GQED_4$ à la Källén}\label{Perturbation}

This section initiates a novel for interacting system with higher order derivative terms. As we already known the behavior of free theory, the next reasonable step relies on the fact that we shall begin to describe the details of the interaction between Podolsky and Fermion particles, while giving a shorter introduction for perturbation approach à la Källén in HP \cite{Kallen1,Kallen2}. A possible general Lagrangian is

\begin{equation}
    \mathcal{L} =  \mathcal{L}_{A} + \mathcal{L}_{\psi} + \mathcal{L}_{I}
\end{equation}

where

\begin{equation}\begin{aligned}\label{puts}
\mathcal{L}_{A} &= -\frac{1}{4}F^{\mu\nu}F_{\mu\nu} + \frac{a^2}{2} \partial^\mu F_{\mu\beta}\partial^\alpha F_{\alpha\beta} - \frac{1}{2} (1 - a^2 \Box) (\partial_\mu A^\mu)^2 \\
\mathcal{L}_{\psi} &=
\Bar{\psi} (\gamma^\mu \partial_\mu + m) \psi  \\
\mathcal{L}_{I} &=   A_\mu  j^{\mu }
\end{aligned}\end{equation}

We observe that Lagrangian of the particle $\mathcal{L}_{\psi}$ and interactions $\mathcal{L}_{I}$ are left unchanged due to Lorentz invariance either Podolsky or Maxwell-Lorentz is used. Applying the standard way, we obtain a couple of differential equations for the field operator in HP

\begin{equation}\label{cde}
\bar{\psi}(x)(\overleftarrow{\partial}_{\mu} \gamma^{\mu} + m ) = ie\bar{\psi}(x) \gamma^{\mu} A_{\mu} (x), \\
\end{equation}

\begin{equation}\label{cde2}
(\gamma^\mu \partial_\mu + m)  \psi(x) = i e \gamma^\mu A_\mu (x) \psi (x), \\
\end{equation}

\begin{equation}\label{cde3}
(1 - a^2 \Box)\Box A^\mu (x) = -\frac{ie}{2} [\Bar{\psi} (x), \gamma^\mu \psi (x) ].
\end{equation}

The exact form for the solutions above can be obtained by solving with the same techniques used in inhomogeneous equations \eqref{wtf1}, \eqref{ihf} and \eqref{wtf2}. We get

\begin{equation}\label{p1}
\psi(x) = \psi^{(0)}(x) - \int d^4 x' S_R (x-x')i e \gamma A(x') \psi(x')  
\end{equation}
\begin{equation}\label{p2}
\bar{\psi}(x) = \bar{\psi}^{(0)}(x) - \int d^4 x' \bar{\psi}(x') i e \gamma A(x') S_A (x'-x)  
\end{equation}
\begin{equation}\label{p3}
A_\mu (x) = A^{(0)}_\mu (x)-\int d^4 x' D_R^{P}(x-x')ie[\Bar{\psi} (x') ,\gamma_\mu \psi(x')]      
\end{equation}

Where $ \{ \psi^{(0)}(x),\bar{\psi}^{(0)}(x),A^{(0)}_\mu(x)\}$ are the initial value at time $x_o = -\infty$ in HP. These covariant equations of motion have the same physical behavior and symmetries of the differential \cref{cde,cde2,cde3}. At low energy, we propose an expansion of field operator in power series as a useful insight to study infrared dynamics since the gauge coupling $\frac{e^2}{4 \pi} \approx \frac{1}{137}$ is small \cite{Kallen2},

\begin{equation}\label{P1}
\psi(x)=\psi^{(0)}(x)+e\psi^{(1)}(x)+e^{2}\psi^{(2)}(x)+ \ldots ,
\end{equation}
\begin{equation}\label{P3}
\bar{\psi}(x)=\bar{\psi}^{(0)}(x)+e\bar{\psi}^{(1)}(x)+e^{2}\bar{\psi}^{(2)}(x)+\ldots ,
\end{equation}
\begin{equation}\label{P2}
A_{\mu}(x)=A_{\mu }^{(0)}(x)+eA_{\mu }^{(1)}(x)+e^{2}A_{\mu }^{(2)}(x)+ \ldots .
\end{equation}

After replacing these expansions into differential field \cref{cde,cde2,cde3}. We shall organize the terms perturbatively in powers of coupling constant, and so obtain the general recursive relations for field operators in the HP

\begin{equation}\label{RRpsi}\begin{aligned}
\psi^{(n+1)}(x)
&= - \frac{i}{2}\int{d^4x'}S_R(x-x')\gamma^\mu\sum_{m=0}^{n}\{A_{\mu }^{(m)}(x'),\psi^{(n-m)}(x')\},\\
\end{aligned}\end{equation}

\begin{equation}\begin{aligned}\label{RRApsi}
\bar{\psi}^{(n+1)}(x)
&=-\frac{i}{2}\int{d^4x'}\sum_{m=0}^{n}\{A_{\mu }^{(m)}(x'),\psi^{(n-m)}(x')\}\gamma^\mu S_A(x'-x),
\end{aligned}\end{equation}

\begin{equation}\label{SolA1gPS1}\begin{aligned}
A_{\mu }^{(n+1)}(x)
&=\frac{i}{2}\int{}d^{4}x' D_P^R (x-x') \sum_{m=0}^{n} [\bar{\psi}^{(m)}(x')\gamma_{\mu},\psi^{(n-m)}(x') ],
\end{aligned}\end{equation}

where $n \geq 0$. For sake of clarity, we carry out a symmetrization procedure in \eqref{RRpsi} and \eqref{RRApsi} since at the equal time the free field operators $A_{\mu }^{(0)}(x)$ and $\psi^{(0)}(x)$ commute. We shall identify the first and second order for the fermion operator $\psi (x)$

\begin{equation}\begin{aligned}\label{111}
\psi^{(1)}(x) &= - i \int d^4x'S_R(x-x') \gamma^\nu A^{(0)}_\nu (x') \psi^{(0)}(x')  \\ 
\psi^{(2)}(x) &= \frac{1}{4}  \int{d^4 x'} \int{d^4 x''} S_R(x-x') \gamma_\mu \left\{ \psi^{(0)}(x'),[\bar{\psi}^{(0)} (x''), \gamma^\mu  \psi^{(0)}(x'')] \right\}  D_P^R (x'-x'') \\
& -\frac{1}{2}  \int{d^4 x'} \int{d^4 x''}    S_R(x-x') \gamma^{\nu_1} S_R (x'-x'') \gamma^{\nu_2} \psi^{(0)} (x'') \{ A^{(0)}_{\nu_1} (x'), A^{(0)}_{\nu_2} (x'')   \},
\end{aligned}\end{equation} 

the conjugate operator $\Bar{\psi} (x)$

\begin{equation}\begin{aligned}\label{333}
\Bar{\psi}^{(1)}(x) &= - i \int d^4x' \bar{\psi}^{(0)}(x') \gamma^\mu A^{(0)}_\mu (x') S_A(x'-x)  \\
\Bar{\psi}^{(2)}(x) &= \frac{1}{4}\int{d^4x'}\int  d^4x'' \Big\{[\bar{\psi}^{(0)}(x''), \gamma_{\mu} \psi^{(0)}(x'')], \bar{\psi}^{(0)}(x') \Big\} \gamma^{\mu}S_A(x'-x) D_P^A (x'-x'') \\ & - \frac{1}{2}\int{d^4x'}\int{d^4 x''}\bar{\psi}^{(0)}(x'') \gamma^{\nu}S_A(x''-x')\gamma^{\mu}S_A(x'-x) \{A_{\nu }^{(0)}(x'),A_{\mu }^{(0)}(x'')\},
\end{aligned}\end{equation}

and the gauge operator $A_{\mu } (x)$ 

\begin{equation}\begin{aligned}\label{222}
A_{\mu }^{(1)}(x) &=  \frac{i}{2} \int{d^4 x'}  D_P^R (x-x') [ \Bar{\psi}^{(0)} (x'), \gamma_\mu   \psi^{(0)}(x') ] \\
A_{\mu }^{(2)}(x) &= \frac{1}{2}\int d^{4}x'\int{d^4x''} D_P^R (x-x') \bigg( [\bar{\psi}^{(0)}(x') ,\gamma_{\mu} S_R(x'-x'')\gamma^\nu \psi^{(0)}(x'') ]  \\ 
& + [\bar{\psi}^{(0)}(x'')\gamma_\nu  S_A(x''-x'),\gamma_{\mu}\psi^{(0)}(x') ] \bigg) A_{\nu }^{(0)}(x'').  
\end{aligned}\end{equation}

Our main goal is to determine the physical prediction of vacuum polarization from the observable current operator. Thereby, let us consider a perturbative expression of physical observable $j^\mu (x)$. In a similar fashion of the above operators, the current is given by

\begin{equation}\begin{aligned}\label{EXPCurren}
j_{\mu} (x) 
&= j^{(0)}_{\mu} (x) + ej^{(1)}_{\mu} (x) + e^2 j^{(2)}_{\mu} (x) + \ldots.
\end{aligned}\end{equation}

It is easy to compute $j^{(n)}_{\mu}(x)$ by a straight replacement of the recursive relations \eqref{RRpsi} and \eqref{RRApsi} in the eq. \eqref{CORRP}. Hence, comparing terms with same order of coupling. We shall introduce the lowest order terms of current operator here. The zero order

\begin{equation}\begin{aligned}\label{j0}
j^{(0)}_{\mu}(x) &= \frac{i}{2}[\Bar{\psi}_0(x), \gamma^\mu \psi_0(x)],
\end{aligned}\end{equation}

the first order

\begin{equation}\begin{aligned}\label{j1}
j^{(1)}_{\mu} (x) &= \frac{1}{2}\int{d^4x'} \bigg( [ \bar{\psi}^{(0)}(x),\gamma_{\mu}{S_R}(x-x') \gamma^{\nu} \psi^{(0)}(x') ] + [ \bar{\psi}^{(0)}(x')\gamma_{\nu}{S_A}(x'-x),\gamma_{\mu}\psi^{(0)}(x) ] \bigg)  A_{\nu}^{(0)}(x'),
\end{aligned}\end{equation}

and the second order

\begin{equation}\begin{aligned}\label{j2}
j^{(2)}_{\mu} (x)
&=  \frac{i}{8}\int{d^4x'}\int{d^4x''} [ \bar{\psi}^{(0)}(x),\gamma_{\mu} S_R(x-x')\gamma_\nu \{ \psi^{(0)}(x') , [ \bar{\psi}^{(0)}(x''), \gamma^\nu \psi^{(0)}(x'') ] \} ]  D_P^R (x'-x'')\\
& \quad - \frac{i}{4}\int{d^4x'}\int{}d^4x'' [ \bar{\psi}^{(0)}(x), \gamma_\mu S_R(x-x') \gamma^{\nu_1} S_R(x'-x'')  \gamma^{\nu_2} \psi^{(0)}(x'') ] \{ A_{\nu_1}^{(0)}(x') , A_{\nu_2 }^{(0)}(x'')   \} \\
&\quad - \frac{i}{2} \int{d^4x'} \int d^4x'' [ \bar{\psi}^{(0)}(x') \gamma^{\nu_1}  A_{\nu_1}^{(0)}(x') S_A(x'-x), \gamma_\mu S_R(x-x'') \gamma^{\nu_2}  A_{\nu_2 }^{(0)}(x'') \psi^{(0)}(x'')   ] \\
&\quad + \frac{i}{8} \int{d^4x'} \int{d^4x''} [ \{ [ \bar{\psi}^{(0)}(x'') , \gamma^{\nu} \psi^{(0)}(x'') ], \psi^{(0)} (x') \} \gamma_\nu  S_A (x'-x), \gamma_\mu \psi^{(0)} (x)] D_P^R (x'-x'') \\
&\quad-\frac{i}{4}\int{d^4x'}\int{d^4x''} [ \bar{\psi}^{(0)}(x'')\gamma^{\nu_2}S_A(x''-x')\gamma^{\nu_1}S_A(x'-x),\gamma_{\mu}\psi^{(0)}(x) ] \{ A_{\nu_1}^{(0)}(x') , A_{\nu_2 }^{(0)}(x'')\}
\end{aligned}\end{equation}

We will discuss further details of the interplay between the current operator and vacuum polarization tensor in the next section.

\section{Vacuum Polarization}\label{VAC}
 
Unlike renormalization methods used widely in ordinary quantum field theory \cite{Akhiezer,Scharf} and expressed in IP, we propose a framework-perturbation model developed until the present moment, which is very convenient for HP. Our main objective is to provide if the Podolsky structure gets rid of difficulty caused by the natural appearance of ultraviolet divergences in $QED_4$. This section is dedicated to analyzing the radiative correction of vacuum polarization in the presence of a small unquantized external field described by the potential $A_{\mu}^{ext}(x)$. Due to linearity of $GQED_4$, we may introduce in the Lagrangian the term $j^\mu A^{ext}_\mu$ in \eqref{puts}, which is a straightforward calculation to the equation of motions

\begin{equation}\begin{aligned}\label{psip1}
\psi(x)&=\psi^{(0)}(x)-e\int{d^4x'}S_R(x-x')\gamma^{\mu}\Big(A_{\mu}(x')+A_{\mu}^{ext}(x')\Big)\psi(x'),
\end{aligned}\end{equation}

\begin{equation}\begin{aligned}\label{psip2}
\bar{\psi}(x)
&=\bar{\psi}^{(0)}(x)-e\int{d^4x'}\bar{\psi}(x')\gamma^{\mu}\Big(A_{\mu}(x')+A_{\mu}^{ext}(x')\Big)S_A(x'-x).\\
\end{aligned}\end{equation}

Where $A_{\mu}^{ext}(x)$ is fixed independent of the particular configuration of the system and its physical value is determined experimentally. Substituting the equations \eqref{psip1} and \eqref{psip2} in the current operator \eqref{EXPCurren}, we obtain the following expression up to first order

\begin{equation}\label{EXPCurrenM}\begin{aligned}
j^{\mu}(x)&=\frac{ei}{2}[\overline{\psi}^{0}(x)\gamma^{\mu},\psi^{0}(x)] \\
& + \frac{e^2}{2}\int{d^4x'}[\bar{\psi}^{(0)}(x),\gamma^{\mu}{S_R}(x-x')\gamma^{\nu}\psi^{(0)}(x')] \big(A_{\nu}^{(0)}(x')+A_{\nu}^{ext}(x')\big)\\
&+ \frac{e^2}{2}\int{d^4x'}[\bar{\psi}^{(0)}(x')\gamma^{\nu}{S_A}(x'-x)\gamma^{\mu},\psi^{(0)}(x)]\Big(A_{\nu}^{(0)}(x')+A_{\nu}^{ext}(x')\Big) \\
&\quad+e^2j^{(2)}_{\mu}(x) + \ldots.
\end{aligned}\end{equation}

It is obvious that the current operator \eqref{EXPCurrenM} has a non-vanishing vacuum expectation value

\begin{equation}\begin{aligned}\label{kubo}
\langle0|j^{\mu}(x)|0\rangle
&=\int{d^4y}K^{\mu\nu}(x-x')A_{\nu}^{ext}(x'),\\
\end{aligned}\end{equation}

As it appears, the polarization tensor $K^{\mu\nu}(x-x')$ describes the proportionality between induced current and $A_{\nu}^{ext}(x')$. In lowest order, we can write following \eqref{EXPCurrenM}

\begin{equation}\label{TensK}\begin{aligned}
K^{\mu\nu}(x-y)
&= \frac{e^2}{2} \bigg( Tr[\gamma^{\mu}S_R(x-y)\gamma^{\nu}S^{(1)}(y-x)] + Tr[\gamma^{\mu} S^{(1)}(x-y) \gamma^{\nu} S_A(y-x)] \bigg).\\
\end{aligned}\end{equation}

We predicted that results of $GQED_4$ are explicitly gauge independent as well as $QED_4$ since both of them have the same symmetries. It is convenient to employ the Fourier transformation

\begin{equation}
K_{\mu\nu}(x-x') = \frac{1}{(2\pi)^4}\int d^4p e^{ip(x-x')} K_{\mu\nu}(p).
\end{equation}

Then the tensor \eqref{TensK} should be

\begin{equation}\begin{aligned}\label{fffddd}
K_{\mu\nu}(p) &=  \frac{1}{(2\pi)^4} \int dp' dp^{''} \delta^4(p-p'+p^{''})Sp[\gamma_\mu (i\gamma p'-m)\gamma_\nu(i\gamma p^{''}-m)] \times  \\
& \bigg( \delta^4(p'+ m^2)( P \frac{1}{p^{2''} + m^2 } -i \pi \epsilon(p^{''}) \delta^4(p^{2''}+ m^2)  )  \\
& + \delta^4(p^{''}+ m^2)( P \frac{1}{p^{2'} + m^2 } + i\pi \epsilon(p') \delta^4(p'+ m^2))\bigg).
\end{aligned}\end{equation}

As vacuum polarization is a second order tensor, the general way to express through irreducible representations becomes

\begin{equation}\label{kmn}
K^{\mu\nu}(p) = A(p^2) p^\mu p^\nu - p^2 g^{\mu\nu} B(p^2)
\end{equation}

where $g^{\mu\nu}$ and $k^\mu$ are Lorentz covariant objects and $A(p^2)$ and $B(p^2)$ are invariant scalar functions. The induced current must be conserved and invariant under gauge symmetry so that we expect $p^\mu K_{\mu\nu}(p)= 0$ in momentum space. One may be tempted to conclude this represent the rightful form of polarization tensor. Nonetheless, the following arguments in \cite{galvao} and the requirement of a covariant photon. The eq. \eqref{kmn} should be written as 
 
\begin{equation}\label{Kmunu}
K_P^{\mu\nu}(p)=[a^2 p^2 p^{\mu}p^{\nu} - a^2 p^4 g^{\mu\nu} + p^{\mu}p^{\nu} -  p^2 g^{\mu\nu}] G(p^2),
\end{equation}

We construct a new physical tensor symmetric and transverse where Podolsky mass emerges naturally. $QED_4$ may suggest the format \eqref{kmn} for polarization tensor in Podolsky framework, for example see \cite{Bufaloart}. However, a careful analysis of \eqref{super} tells us we can split the polarization tensor in a superposition contribution through the new degree of freedom to ensure a transverse photon. Note that the tensor ends up $ \lim_{a \to 0 } K_P^{\mu\nu}(p)= K^{\mu\nu}(p)$ in a similar way that Podolsky reduces to Maxwell.

An interesting consequence of \eqref{kubo} is that $e^2 K^{\mu\nu} (p) A_\nu (p)$ relates the Fourier transform of induced current via external potential in vacuum medium, then it is evident that external current is also unchanged under gauge transformation. This line of thought is sufficient to demand the appearance of tensor $K_P^{\mu\nu}$ not only as virtual electron-positron amplitude but also can be understood as integral equivalent to external photon dynamic. In such a way, the external potential is connected with external current by dynamical equation \cite{Bogoliubov}. Starting from \eqref{Kmunu} and considering the Kubo formula, we shall obtain the induced current

\begin{equation}\begin{aligned}
\langle 0 | j_{\mu}(x) | 0 \rangle 
&= \frac{1}{(2 \pi)^4} \int dp dx' e^{ip(x-x')}G(p) [  \Box A_\mu^{(ext)} - \partial_\mu \partial^\nu A_\nu^{(ext)}(x') + a^2 \Box^2 A_\mu^{(ext)}(x') \\
& - a^2 \Box \partial_\mu \partial^\nu A_\nu^{(ext)}(x') ] \\
&= - \frac{1}{(2 \pi)^4} \int dp dx' e^{ip(x-x')}G(p) [ j^{(ext)}_\mu (x')] 
\end{aligned}\end{equation}

This explicit assumption, with the limit $a \rightarrow 0$, is used in Ref. \cite{Kallen2}. Hence, the vacuum polarization effects including the characteristic length of Podolsky and the higher derivative terms. The vacuum-to-vacuum transition amplitude still depends on a linear external current and we use the contracted form of \eqref{fffddd} to better evaluate the polarization tensor. Thus, we have

\begin{equation}\begin{aligned}\label{rty1}
G(p) 
&= -\frac{K^{\mu}_{\mu} (p)}{3p^2(1 + a^2 p^2)} = \frac{1}{48 \pi^3 p^2 (1 + a^2 p^2)} \int dp' dp^{''} \delta^4(p-p'+p^{''})\times \\
& Sp[\gamma_\mu (i\gamma p'-m)\gamma_\mu(i\gamma p^{''}-m)]   \bigg\{ \delta^4(p^{2'}+ m^2) \bigg( \mathcal{P} \frac{1}{p^{2''} + m^2 } -i\pi \epsilon(p^{''}) \delta^4(p{''}+ m^2)  \bigg) \\ 
& +   \delta^4(p^{''}+ m^2)\bigg( \mathcal{P} \frac{1}{p^{2''} + m^2 } + i\pi \epsilon(p') \delta^4(p'+ m^2)  \bigg)    \bigg\}
\end{aligned}\end{equation}

A new observable emerges from the sum of classical external and induced current

\begin{equation}\label{jobs0}\begin{aligned}
j_{obs}^\mu (x)
&= j_{ext}^{\mu}(x)+\langle0|j^{\mu}(x)|0\rangle
\end{aligned}\end{equation}

where

\begin{equation}\label{jobs01}\begin{aligned}
j_{obs}^\mu (x) &=\frac{1}{(2\pi)^4} \int d^4p d^4x' [1 - G(p)] j^\mu_{ext} (x') e^{ip(x-x')}.\\
\end{aligned}\end{equation}

One motivation to study this subject is the possibility to analyze the space-time structure of an interacting theory in the presence of a classical external flux $j^\mu_{ext} (x)$. Moreover, the averaging value of $j_{obs}^\mu (x)$ is not a direct observable because the properties of the vacuum cannot be measurable on a single momentum scale. In other words, this process is equivalent to a redefinition of electrical charge even if the $G(p)$ is dependent on scale momentum. One can solve by fixing the value of $j_{obs}^\mu(x)$ at a particular momentum scale ("frequency") known experimentally and so relating to $G(p)$. Thus, the starting point may be a suitable approximation to the infrared regions where the system minimizes the influence of quantum fluctuation. Therefore, we can assure the equivalence between external and observed current $j^\mu_{obs}(x) = j^\mu_{ext}(x)$ by a slow virtual photon ($p \approx 0$) (classical Coulomb scattering). Since \eqref{rty1} has a non-vanishing value at null momentum, it is not possible to set $G(0)=0$. Consequently, the renormalization of charge for the observable current becomes

\begin{equation}\label{jobs1}\begin{aligned}
 j_{obs}^\mu (x)  =  \frac{1}{(2\pi)^4} \int d^4p d^4x' [1 - G(p) + G(0)] j_{ext}^\mu (x') e^{ip(x-x')}.\\
\end{aligned}\end{equation}

this expression is a linear response due to disturbance of external current operator where $G(0)$ plays the role of renormalized charge. It is natural to rewrite the polarization tensor following \eqref{kubo} and \eqref{jobs0}

\begin{equation} 
K^{\mu\nu}(p) =  (1 + a^2 p^2)(p^\mu p^\nu - g^{\mu\nu} p^2 ) [ G(p^2) - G(0) ]
\end{equation}

For the case of polarization tensor, we keep track the quantity $G(p^2)$. Firstly, we shall evaluate imaginary part of \eqref{rty1}. Thus

\begin{equation}\begin{aligned}\label{rty}
& Im G(p) = -\frac{K_{\mu}^{\mu}(p)}{(1 + a^2 p^2)3p^2}= \frac{1}{ (1 + a^2 p^2) 6 \pi^2 p^2} \int d^4p' d^4p^{''} \delta^4(p-p'+p^{''})Sp[  p'(p'-p) + 2 m^2] \\
& \times \bigg\{\delta^4(p^{2'}+ m^2)(\epsilon(p^{''}) \delta^4(p{''}+ m^2))-\delta^4(p^{''}+ m^2)( \epsilon(p') \delta^4(p'+ m^2)) \bigg\}
\end{aligned}\end{equation}

integrating into the rest frame $\Vec{p}=0$, and after that generalizing by a boost $ p^2_o \rightarrow - p^2 $. Finally, we find

\begin{equation}\label{TFG29}\begin{aligned}
ImG(p)= \pi \epsilon(p)\Pi^{(0)}(p^2)  
\end{aligned}\end{equation}

where

\begin{equation}\label{PIso}
\Pi^{(0)}(p^2)= \frac{e^2}{12 \pi^2 (1 + a^2 p^2)}(1 - \frac{2m^2}{p^2})  \sqrt{1 + \frac{4m^2}{p^2}} \Theta(\frac{-p^2}{4} - m)
\end{equation}

The imaginary part vanishes in case of $p^2 < 4 m$. A Hilbert transformation of analytical function $G(p)$ illustrates an interplay between the imaginary and real parts and encodes the causality principle.

\begin{equation}
Re G(\Vec{p},p_o) = \frac{1}{\pi} \mathcal{P} \int^{+\infty}_{-\infty}
 \frac{Im G(\Vec{p},x)}{x-p_o} dx
\end{equation}

\begin{equation} 
 Im G(\Vec{p},p_o) = - \frac{1}{\pi} \mathcal{P} \int^{+\infty}_{-\infty}
 \frac{Re G(\Vec{p},x)}{x-p_o} dx
\end{equation}

Now we can derive the real part

\begin{equation}\begin{aligned}\label{dede}
Re G(\Vec{p},p_o) &= \mathcal{P} \int^{+\infty}_{-\infty} dx \frac{x}{|x|} \frac{\bar{\Pi}^{(0)}(\Vec{p}^2 - x^2)}{(x-p_o)} = \mathcal{P} \int^{+\infty}_0  dx\bar{\Pi}^{(0)}(\Vec{p}^2 - x^2)[\frac{1}{(x-p_o)} + \frac{1}{(x + p_o)}]  \\
& = \mathcal{P} \int^{+\infty}_0 \bar{\Pi}^{(0)}(\Vec{p}^2 - x^2)[\frac{2x dx}{(x^2-p^2_o)}]  = \mathcal{P} \int^{+\infty}_{4m^2} \frac{\bar{\Pi}^{(0)}(-s)}{(s + p^2)}ds \equiv \bar{\Pi}^{(0)}(p^2).
\end{aligned}\end{equation}

The Podolsky's parameter has no direct influence over the integration limit. One finds

\begin{equation}\label{por}
G(p) = \bar{\Pi}^{(0)}(p^2) + i \pi \epsilon \Pi^{(0)}(p^2)
\end{equation}

It is time to notice that the gauge dependence $\xi$ drops out from this physical process. We observe the same logarithmically divergent at $QED_4$. Then the Podolsky framework can no longer improve the convergence of integral amplitude as hoped for higher derivative Lagrangians and we found this same conclusion in IP, see Ref. \cite{bufalogauge}. A very useful procedure of renormalization to get a finite polarization tensor is including $G(0)$  by a subtraction process in \eqref{por}.

\begin{equation}
   G(p) - G(0) = \bar{\Pi}^{(0)}(p^2) - \bar{\Pi}^{(0)}(0) + i \pi \epsilon \Pi^{(0)}(p^2) 
\end{equation}

the physical meaning of subtraction dispersion integral with different energy scale is the renormalization of charge electric. Thus, the form of polarization tensor is

\begin{equation}\label{kkk}
K_{\mu\nu} =  [ a^2 p^4 g_{\mu\nu} - a^2 p^2 p_\mu p_\nu +  p^2 g_{\mu\nu} - p_\mu p_\nu ] \frac{e^2}{12 \pi^2} \int^{+\infty}_{4m^2}ds  \frac{(1 + \frac{2m^2}{s})  \sqrt{1 - \frac{4m^2}{s}}}{(1 - a^2 s )s(s + p^2)}      
\end{equation}

$GQED_4$ yields a correction on vacuum dependent on the length of Podolsky. If $a \rightarrow 0$, we shall recover the vacuum polarization tensor of $QED_4$ \cite{Bogoliubov}. Instead of electric charge, $m_P$ has no signature on the renormalization of theory. However, it plays a role in Coulomb scattering and running coupling \cite{Bufaloart}. Finally, after some basic manipulations, the exact solution of integral above reads

\begin{equation}\begin{aligned}\label{1234}
& \bar{\Pi}^{(0)}(p^2) - \bar{\Pi}^{(0)}(0) = \frac{e^2}{12 \pi^2} \bigg[ \frac{2}{3} - \frac{p^2}{2 m^2}  -  \frac{4 m^2}{m_P^2} + \frac{1}{2} \frac{ \bigg( 1 - \frac{p^2}{m_P^2} \bigg) \bigg( 1 - \frac{2 m^2}{p^2}  \bigg)  \bigg( 4 + \frac{p^2}{m^2} \bigg)}{1 + p^2/m_P^2 }  - \bigg( 1 - \frac{2 m^2}{p^2} \bigg) \times \\ 
& \sqrt{1+\frac{4m^2}{p^2}} \ ln \frac{1 + \sqrt{1 + 4m^2/p^2}}{|1 - \sqrt{1 + 4m^2/p^2}|} +  \bigg( \frac{p^2}{4 m^2} + \frac{p^2}{m_P^2} \ ln  \frac{4m^2}{4m^2 + m_P^2} \bigg) \bigg( 1  - \frac{2 m^2}{p^2} \bigg) \bigg[ \bigg(\frac{   1 -  \frac{4m^2}{m_P^2}   }{1 + p^2/m_P^2} \bigg)  \times \\ 
& \bigg[ 1  + \bigg( 1 - \frac{4m^2}{p^2} \bigg) \frac{p^2}{m_P^2} \bigg] - 1  \bigg]  \bigg] 
\end{aligned}\end{equation}

If $1 + \frac{4m^2}{p^2} < 0$ in the equation \eqref{kkk}, we must have arctan function rather than logarithm. In principle, It would be interesting to analyze the physical effect of two different energy regimes. The small momentum and a large one. For small values of $|\frac{p^2}{m^2}|$ the leading order of the energy-shift reads

\begin{equation}\label{123}
\Delta V^P (p) = V^P (p) - V^P_o (p) = \frac{e^2}{p^2} (\bar{\Pi}^{(0)}(p^2) - \bar{\Pi}^{(0)}(0))  =  - \frac{e^4}{60 \pi^2 m^2} + \frac{5 e^4}{6 \pi^2 m_P^2} + \ldots ,
\end{equation}

where $V^P_o (p) = e^2 ( 1/p^2 - 1/(p^2 + m_P^2 ) ) $ is the classical Podolsky potential in Fourier space \cite{Podolskyorigin}. We call attention to the fact that Podolsky information over photon self-energy is directly associated with the external photon and gauge coupling $(e)$ from the closed fermion loop \eqref{TensK}, thus, by the consideration applied so far it should be expected that Podolsky fails to give a proper convergence on photon propagator when only the fermions are involved in radiative correction. The influence of $GQED_4$ is only felt in quantum fluctuations of the vertex part and fermionic propagator.

Furthermore, a careful analysis shows that we can split up the eq. \eqref{123}, in the same way as \eqref{super}, into $QED_4$ and $GQED_4$ contribution from the first and second terms, respectively. We may wonder what is the physical meaning of a positive signal from Podolsky mass term. One might suspect if we look at the electrostatic potential of Podolsky \cite{Podolskyorigin} and thus identifying that the $m_P$ term acts as a "screening" process. It is expected that this contribution at low-energy is significantly smaller than the leading $QED_4$ term. Otherwise, the effect would be detected. Besides that, we calculate the influence of $GQED_4$ correction by the following equation.

\begin{equation}
    \delta = \bigg( \Delta V (p) \bigg)_{GQED}  \bigg/ \bigg( \Delta V (p) \bigg)_{QED} - 1
\end{equation}

In this case, we use the experimental data of electron anomalous magnetic momentum $m_P \geq 37.59 GeV/c^2 $ in Ref. \cite{Bufaloart} and the electron mass $m = 0.511 MeV/c^2 $ to compute eq. \eqref{123}.  We can roughly evaluate the leading contribution to this process $ \delta = - 0.923 \times 10^{-5} \ \% $. A reasonable value, otherwise, the experiments would have been detected the deviations from $QED_4$ a long time. By the way, in low-energy scenario, one can argue the $GQED_4$ and $QED_4$ are the same theory when radiative corrections with higher order than Podolsky contribution, as for instance spin-orbit coupling, are negligible in ordinary atom. Hence, we may restore the well known Uehling potential \cite{Bogoliubov}.

The result above was considerably smaller than Uehling potential and showed that Podolsky would be more visible at the higher energy. The new parameter of $GQED_4$ should allow us to write a new perturbation scheme where the theory continues well defined. For practical reason $ m^2 \leq p^2 < m_P^2 $ and the potential of equation \eqref{1234} reads

\begin{equation}\label{12345}
\Delta V (p) = V(p) - V_o (p) =  \frac{e^4}{12 \pi^2 p^2} \  ln \bigg( \frac{p^2}{m^2} \bigg) - \frac{e^4}{12 \pi^2} \frac{1}{m_P^2} \  ln \bigg( \frac{p^2}{m^2} \bigg) + \mathcal{O} \  ln \bigg( \bigg( \frac{p^2}{m_P^2} \bigg) \bigg) + \ldots 
\end{equation}

The expansion must be done with proper care since the regime has two important constant: electron and Podolsky mass. We notice in the small distance the logarithmic turns out important. Furthermore, we expect further results present in many efforts to study experimentally an upper limit of $GQED_4$: ion interferometer \cite{probePodolsky}, Bhabha electron-positron scattering \cite{Causalapproach}, analysis of cosmic microwave background data \cite{Podolskycosmo}, and Stefan-Boltzmann law \cite{Bonin}. Analyzing the running coupling constant in HP picture will be done in a forthcoming work.

$GQED_4$ may give a contribution towards the direction of solving the discrepancy found in measuring the Lamb shift ($1-S$ ground state) \cite{hydro}, muonic Lamb shift and muon anomalous magnetic moment \cite{bom}, and higher-order polarization at two loops calculated by Källén and Sabry \cite{pqp}. These observations of incorrect compatibility between theory and experiment require careful analysis.



\section{Conclusions and Perspectives}\label{conclu}

In this paper, we have studied vacuum polarization in the generalized quantum electrodynamics following a covariant perturbative framework "à la" Källén in the HP. Higher derivative terms provided wealth information on underlying properties of theory and cannot be treated in the same footing as lower-order Lagrangians. Since stability and unitarity were proven, $GQED_4$ could reap the benefits of does not suffer from the pathology of Ostrogradski.

Once we had identified that induced current was related to external fields linearly by polarization tensor, we investigated Podolsky by means of the same analytical process of $QED_4$. Since the symmetries and the order of divergence were the same for both of them, It was suitable to address the basic idea of charge renormalization. By physical argument, we showed that a new parameter encoded by length $1/m_P$ must be present in vacuum polarization tensor which claimed satisfy the higher order Lagrangian. For renormalization, we had chosen a dispersion relation approach in which the operation of subtracting the dispersion integral with different scale of momentum conducted to fixing unit of charge and convergent polarization.

We wanted to show that Podolsky theory could open a new window to calculate the deviation of $QED$ when all symmetries are still preserved and the analysis performed here was easily implemented in numerical algorithms. Even though the precision to measuring the influence of Podolsky term is not sensible to the experimental error at low energy, the higher energy experiments as $e^- e^+$ electron-positron scattering and proton-proton collision offered a new insight 

As a natural perspective, the continuation of this work will extend the treatment given here to divergence of electron self-energy and the vertex functions in the one-loop approximation where the influence of Podolsky massive photon should be expected to subtract the standard divergent terms from the massless photon of $QED_4$. It should be worth to compare with the same result already obtained in the Interaction picture.

 
%

\section*{acknowledgements}
D. Montenegro thanks to CAPES for full support and B. M. Pimentel for useful conservation. 



\end{document}